\documentclass[aps,prl,floats,twocolumn,showpacs,nobibnotes,
  showkeys,superscriptaddress,reprint]{revtex4-1}

\usepackage[british]{babel}
\usepackage[latin1]{inputenc}
\usepackage{graphicx,epsfig}
\usepackage{graphics,dcolumn,bm,epic, eepic,fleqn,float}
\usepackage{amssymb,amsmath,amsfonts,multirow,rotate,color}
\usepackage{soul}
\usepackage{xcolor}
\usepackage{colortbl}
\usepackage{epstopdf}

\begin{document}

\title{Phase transition in the economically modeled growth of a cellular nervous system}

\author{Vincenzo Nicosia} 
\thanks{V.N. and P.E.V. have equally contributed to this work}
\affiliation{School of Mathematical Sciences, Queen Mary University of
  London, London E1 4NS, United Kingdom}

\author{Petra E. Vertes} \thanks{V.N. and P.E.V. have equally
  contributed to this work} \affiliation{Department of Psychiatry,
  Behavioural and Clinical Neuroscience Institute, University of
  Cambridge, Cambridge CB2 0SZ, United Kingdom}

\author{William R. Schafer} \affiliation{ Medical Research Council
  Laboratory of Molecular Biology, Cambridge CB2 0QH, United Kingdom}

\author{Vito Latora} \affiliation{School of Mathematical Sciences,
  Queen Mary University of London, London E1 4NS, United Kingdom}
\affiliation{Dipartimento di Fisica e Astronomia \& INFN \& Laboratorio
  sui Sistemi Complessi, Universit\`a di Catania, Via S. Sofia 61,
  Catania, Italy}

\author{Edward T. Bullmore}
\thanks{To whom correspondence should be
  addressed. Email:~etb23@cam.ac.uk}
\affiliation{Department of Psychiatry,
  Behavioural and Clinical Neuroscience Institute, University of
  Cambridge, Cambridge CB2 0SZ, United Kingdom}

\affiliation{Cambridgeshire and Peterborough National Health Service
  Foundation Trust, Cambridge CB21 5EF, United Kingdom}

\affiliation{GlaxoSmithKline Clinical Unit Cambridge, Addenbrookes
  Hospital, Cambridge CB2 0QQ, United Kingdom}

\begin{abstract}
Spatially-embedded complex networks, such as nervous systems, the
Internet and transportation networks, generally have non-trivial
topological patterns of connections combined with nearly minimal
wiring costs.  However the growth rules shaping these economical
trade-offs between cost and topology are not well understood.  Here we
study the cellular nervous system of the nematode worm
\textit{C. elegans}, together with information on the birth times of
neurons and on their spatial locations. We find that the growth of
this network undergoes a transition from an accelerated to a constant
increase in the number of links (synaptic connections) as a function
of the number of nodes (neurons). The time of this phase transition
coincides closely with the observed moment of hatching, when
development switches metamorphically from oval to larval stages.  We
use graph analysis and generative modelling to show that the
transition between different growth regimes, as well as its
coincidence with the moment of hatching, can be explained by a dynamic
economical model which incorporates a trade-off between topology and
cost that is continuously negotiated and re-negotiated over
developmental time. As the body of the animal progressively elongates,
the cost of longer distance connections is increasingly
penalised. This growth process regenerates many aspects of the adult
nervous system's organization, including the neuronal membership of
anatomically pre-defined ganglia.  We expect that similar economical
principles can be found in the development of other biological or
man-made spatially-embedded complex systems.
\end{abstract}

\keywords{connectome | generative model | neurodevelopment | spatial
  networks | \textit{C. elegans} }

\maketitle

In the last decade or so there has been an abundance of
studies demonstrating that superficially diverse systems share
important statistical
properties~\cite{albert02,boccaletti06,bullmore09,barabasi12}. Movie
co-star networks, transport and communication systems, gene-gene
interactomes, and many other natural and man-made systems have
similarly complex topological features: they are generally efficient,
small-world, modular systems with a greater-than-random probability of
highly connected nodes or hubs. Many but not all of these
topologically complex systems are also spatially
embedded~\cite{barthelemy11}. For example, both the Internet and the
World Wide Web have non-trivial topologies but only the Internet is
physically instantiated as a network in a metric space. Spatially
embedded networks generally increase in cost with increasing distance
of connections between nodes; and this cost constraint must be
traded-off against the functional advantages of topological features
like hub nodes, robustness, and high global efficiency, that may add
value but at greater than minimal cost~\cite{barthelemy03}.  Nervous
systems share these general economical properties~\cite{bullmore12}:
at all scales of space and time and in all species it is likely that
brain networks are both parsimoniously wired~\cite{chen06} and
topologically complex~\cite{bullmore09}. 

This was first demonstrated in the case of the network of neurons that
comprises the nervous system of the nematode worm, {\em Caenorhabditis
  elegans}~\cite{watts98,latora03}. The brain of the hermaphrodite
worm consists of $279$ neurons (excluding the pharyngeal neurons), and
is a sparse network ($4\%$ of maximum connection density), with the
majority of connections being between cells separated by short
distances ($<10\%$ of the overall body length of the adult worm).
Both sparse connection density and low connection distance are as
expected by the operation of a parsimonious drive to minimize wiring
cost. However, the wiring cost of the {\em C. elegans} connectome is
not strictly minimized~\cite{perez07,perez09,sporns11}: further
reductions of connection distance can be achieved by re-wiring the
biological network {\em in silico}; but only at the expense of
increasing the shortest topological path between
neurons~\cite{kaiser06}, thus reducing the overall system
efficiency. To put it another way, it seems there is a trade-off
between connection distance and topological efficiency in the
organization of the adult nematode worm's nervous system. Topological
efficiency is theoretically advantageous for globally integrated
information processing and coordinated behaviors, but it is
disproportionately expensive to
engineer~\cite{bullmore12,towlson13}. It is arguable that such
economical trade-offs between topological value and physical cost are
likely to be a general selection pressure on formation of spatially
embedded and topologically complex networks. More specifically, we
predicted that economical principles applied dynamically over the
course of developmental time (100s of mins after fertilization) could
provide a reasonable account of the emergence of multiple observed
features of the growth and adult configuration of the nematode's
nervous system.

\section*{Results}
\begin{figure*}[t]
  \begin{center}
    \includegraphics[width=0.95\textwidth,natwidth=17.8cm,natheight=10.19cm]{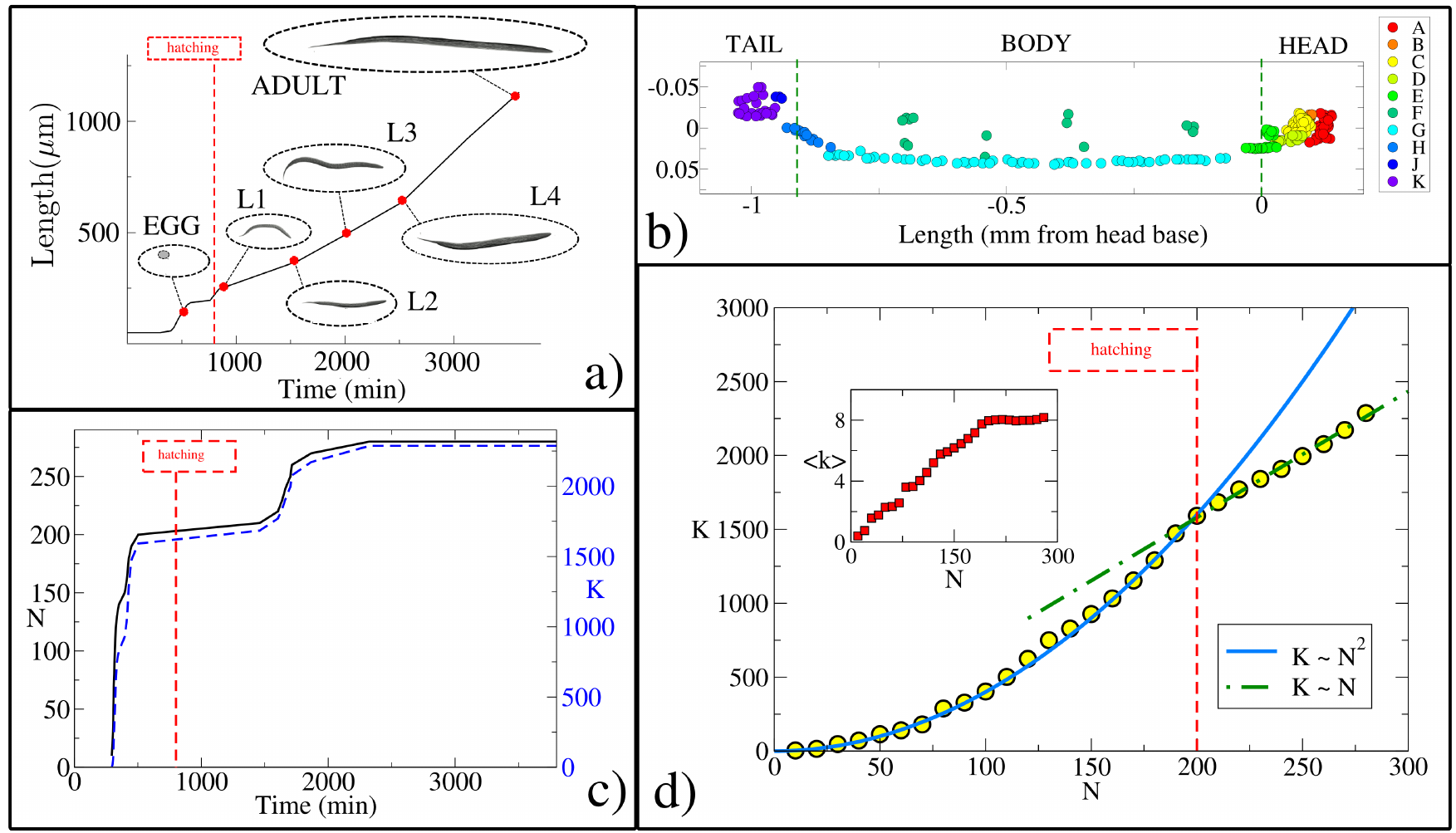}
  \end{center}
  \caption{\footnotesize\textbf{Development of the {\em C.\ elegans}
      nervous system}. \textbf{a)} {\em C.\ elegans} reaches maturity
    roughly 63 hours after fertilization. During this time, its
    body-length increases from $50\mu m$ to $~1130\mu
    m$~\cite{mckeown98, altun05, byerly76}. \textbf{b)} In the adult
    hermaphrodite worm, more than $60\%$ of the neurons are located in
    the head and about $15\%$ are found in the tip of the tail (based
    on data modified according to~\cite{kaiser11}, axis arbitrarily
    centred such that the origin is at the base of the head). Neurons
    are colored by ganglion membership~\cite{varshney11}:
    anterior~[A], dorsal~[B], lateral~[C], ventral~[D],
    retrovesicular~[E], posterior lateral~[F], ventral cord~[G],
    pre-anal~[H], dorso-rectal~[J], lumbar~[K]. \textbf{c)} The total
    number of neurons ($N$, solid black), and connections ($K$, dashed
    blue), grows rapidly between $250$ and $500$ minutes after
    fertilization. Another burst of neurogenesis is observed at the
    end of the L1 larval stage (using data
    from~\cite{kaiser11}). \textbf{d)} Plotting the number of synapses
    as a function of the number of neurons (yellow circles) reveals
    the presence of a phase transition. Before hatching, $K$ grows as
    $N^2$ (solid blue line), while after hatching $K$ grows linearly
    with $N$ (dashed green line). The inset shows the plot of the
    average nodal degree versus $N$.}
  \label{fig:fig1}
\end{figure*}

Here we investigate the growth of the {\em C. elegans} connectome,
from the moment of fertilization through hatching of the egg and
larval elongation to
adulthood~\cite{varshney11,kaiser11}. Importantly, we note that the
physical distances between neurons increase as a function of the
increasing overall length of the worm's body as it matures; see
Figure~\ref{fig:fig1}a.  The cells of the adult nervous system are
concentrated in the head and the tail of the worm, with a series of
neurons running along the length of the body to innervate local muscle
groups (the ventral cord). This system can be decomposed into $10$
ganglia (or neuronal groups) based on anatomical
properties~\cite{arenas08,arenas09}; see Figure~\ref{fig:fig1}b. The
birth times of each neuron tend to cluster in two time windows,
separated by a ``quiet'' period which includes the time of hatching
($800$ minutes after fertilization); see Figure~\ref{fig:fig1}c.  The
developmental changes in the number of nodes ($N$) and edges ($K$) in
the network occur in the context of progressive elongation of the
worm's body, from less than $50\mu m$ before hatching to more than $1
mm$ in the adult.

The two growth spurts in neuronal number, before and after hatching,
are paralleled by a roughly synchronous increase in the total number
of synaptic connections between neurons (Figure~\ref{fig:fig1}c).
However, the form of the relationship between $N$ and $K$ is evidently
different before and after hatching, as shown in
Figure~\ref{fig:fig1}d. The initial increase in $K$ is well described
by a \textit{quadratic} function of $N$, implying that the average
node degree increases linearly as the network grows (see inset). Then,
at $N\simeq 200$, hatching takes place, marking the metamorphic change
of the worm from egg to larva.
This event coincides with a discontinuous change in growth rules:
after hatching, $K$ increases \textit{linearly} with $N$, so that the
average node degree remains constant. This experimental evidence
suggests that sharp qualitative changes can indeed affect the growth
rules governing the development and the formation of complex
networks~\cite{barabasi99,albert02,boccaletti06}. In this case, the
transition from one growth regime to another coincides with a
metamorphic change of the worm, from egg to larva.

\begin{figure*}[t]
  \begin{center}
    \includegraphics[width=0.95\textwidth,natwidth=17.8cm,natheight=4.19cm]{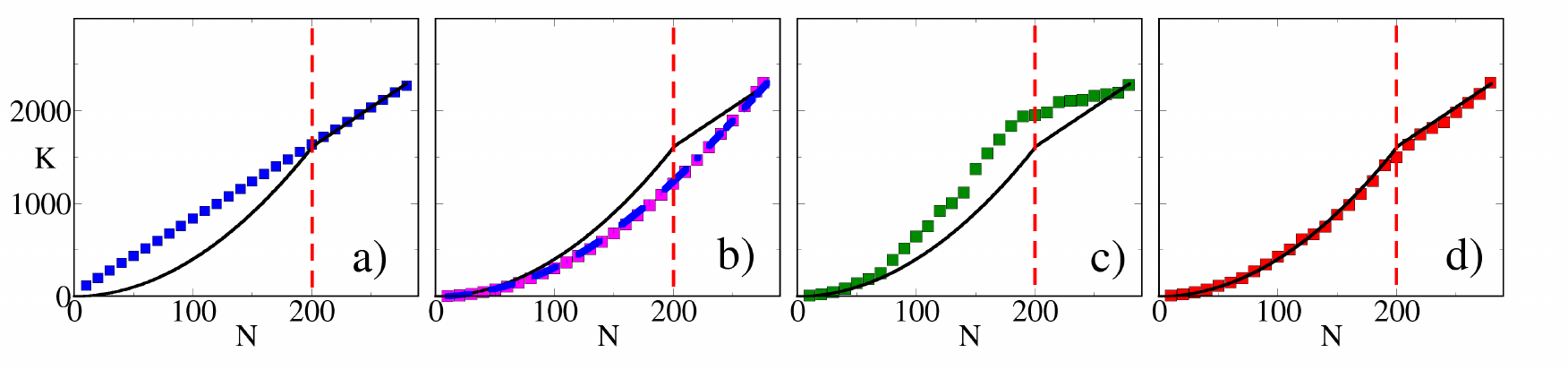}
  \end{center}
  \caption{\footnotesize\textbf{Modeling network growth}.
    \textbf{a)}. The linear preferential attachment model (BA, blue
    squares) fails to reproduce the biphasic growth observed (solid
    line). \textbf{b)}. In the binomial accelerated growth model (BAG;
    magenta squares) and the hidden--variable accelerated growth model
    (HAG; dashed blue line), the average node degree increases
    linearly with the size of the network. \textbf{c)}. The economical
    spatial growth model (ESG; green squares) exhibits a biphasic
    behavior, yielding a transition from quadratic to nearly linear
    growth at $N\lesssim 180$, but fails to capture the details of the
    observed growth. \textbf{d)}. The economical spatio--temporal
    growth model (ESTG; red squares) accurately reproduces the details
    of the biphasic growth trajectory; for example, the inflection
    point of the modeled developmental curve corresponds closely to
    the moment of metamorphosis (hatching). The red dashed line in
    each panel indicates the number of nodes at the time of hatching
    ($N\simeq 200$). The standard error of each growth curve is smaller than
    the size of the symbols used to plot it, and is not reported.}
  \label{fig:fig2}
\end{figure*}

While it is tempting to assume that it is a biological ``trigger'' or
discontinuity associated with hatching that underlies the emergence of
this biphasic growth curve, here we have assessed the ability of
several simple and continuous models of network formation to reproduce
this observed behavior without incorporating further biological
detail; see Figure~\ref{fig:fig2} and Methods. We deliberately decided
to restrict ourselves to stochastic one-parameter models. Firstly,
because our aim was to isolate the fundamental ingredients which could
be responsible for the observed discontinuous growth; secondly
because, as we show in the following, a one-parameter model was indeed
enough to reproduce both the biphasic growth and many of the
structural properties of the adult {\em C. elegans} neuronal network.

The first model we considered was the linear preferential attachment
model, introduced by Barab\'asi \& Albert (BA) ~\cite{barabasi99},
which has been successfully employed to describe the development of
many different complex networks, from the World Wide Web to the
Internet and citation networks. The BA model assumes that the growth
of a network is driven only by its topological properties:
specifically, newborn neurons are more likely to form connections to
neurons that are already well connected. This model predicts a linear
relationship between $N$ and $K$, which matches closely the
post-hatching phase of worm brain development but does not provide a
satisfactory fit to the pre-hatching phase. Conversely, the binomial
accelerated growth (BAG) model, which assumes that the probability of
a connection between a new neuron and any pre-existing neuron is
constant, predicts that $K$ increases as a quadratic function of
$N$~\cite{mendes01}. Similarly, we observe a quadratic dependence of
$K$ on $N$ also in a modified version of accelerated growth (HAG),
which additionally reproduces the node degree distribution of the
adult worm. Accelerated growth models are thus able to reproduce the
pre-hatching phase of the worm brain's growth but fail to accommodate
the transition to linear scaling of $K$ with $N$ in the post-hatching
phase.

We found that economical trade-off models, that take into account the
spatial location of neurons while allowing for some long distance
connections to high degree nodes, were able to reproduce biphasic
growth more accurately. As a first approximation, we defined the
Economical Spatial Growth (ESG) model, which assumes that the
probability of a connection forming between newborn neuron $i$ and
pre-existing neuron $j$ is a product of the degree of the $j$th node
in the adult nervous system, and a decreasing exponential function of
the Euclidean distance $d_{i,j}^{\text{(ad)}}$ between nodes $i$ and
$j$ in the adult worm. Although the modeled growth exhibits two phases, the
transition between quadratic and linear phases occurs before hatching.
Therefore, we considered a more refined Economical Spatio-Temporal
Growth model (ESTG), where $d_{i,j}$ is estimated by the Euclidean
distance between neurons $i$ and $j$ at the time of birth of the
newborn neuron, thereby adjusting for the fact that the connection
distance between any pair of neurons will be shorter at earlier stages
of development before the worm becomes elongated. We extrapolated the
position of each neuron during growth from its position in the adult
worm, assuming that each neuron's position was shifted along the
longitudinal axis in proportion to the overall changes in body length
(see Figure 1a) which we collated from the literature~\cite{mckeown98}
(for the pre-hatching stage) and~\cite{altun05} (after hatching),
using a linear interpolation between larval stages~\cite{byerly76}.
While the penalty on connection distance remains fixed in this model,
its effect on connectivity as a function of the overall scaling of the
system is dynamically evolving. Indeed, the trade-off between distance
and topological degree is increasingly biased in favor of minimizing
connection distance as development proceeds and the worm becomes
longer overall. The model provides an excellent fit to the two
observed scalings of $K$ as function of $N$ in the biological data,
including a good approximation of the moment of hatching to the
transition point from one growth regime to the other.

This suggests that the discontinuity in the growth curve is not
explained by biological triggers related to hatching but is instead a
consequence of the spatial properties of the system. In particular,
the average distance of newly born neurons relative to all other
neurons is much greater after hatching, so that the distance penalty
term begins to dominate the trade-off embodied in the spatial growth
rules. This is especially obvious in the ESTG model, where the worm's
elongation causes distances to increase in the interim between the two
bursts of neurogenesis. Note, however, that a transition is already
visible in the ESG model. This can be explained by noting that most
neurons born after hatching are located along the body of the worm
rather than in the head (see Appendix Section~S1 and Fig.~S1 ), so
that the average distance between these newly born neurons and all
others is again increased after hatching.  We have also tested the
ability of other one-parameter models to reproduce the observed growth
curve (see Appendix Section~S2); in particular, we tried to encode
the cost of long connections through a power-law decay instead of an
exponential one, but none of the alternative models was able to
accommodate the abrupt change in the functional relation between $K$
and $N$ with the same accuracy obtained by ESTG (see  Appendix
Section~S4, Table~S-II and Fig.~S2).

\begin{figure*}[t]
  \begin{center}
    \includegraphics[width=0.95\textwidth,natwidth=17.8cm,natheight=10.2cm]{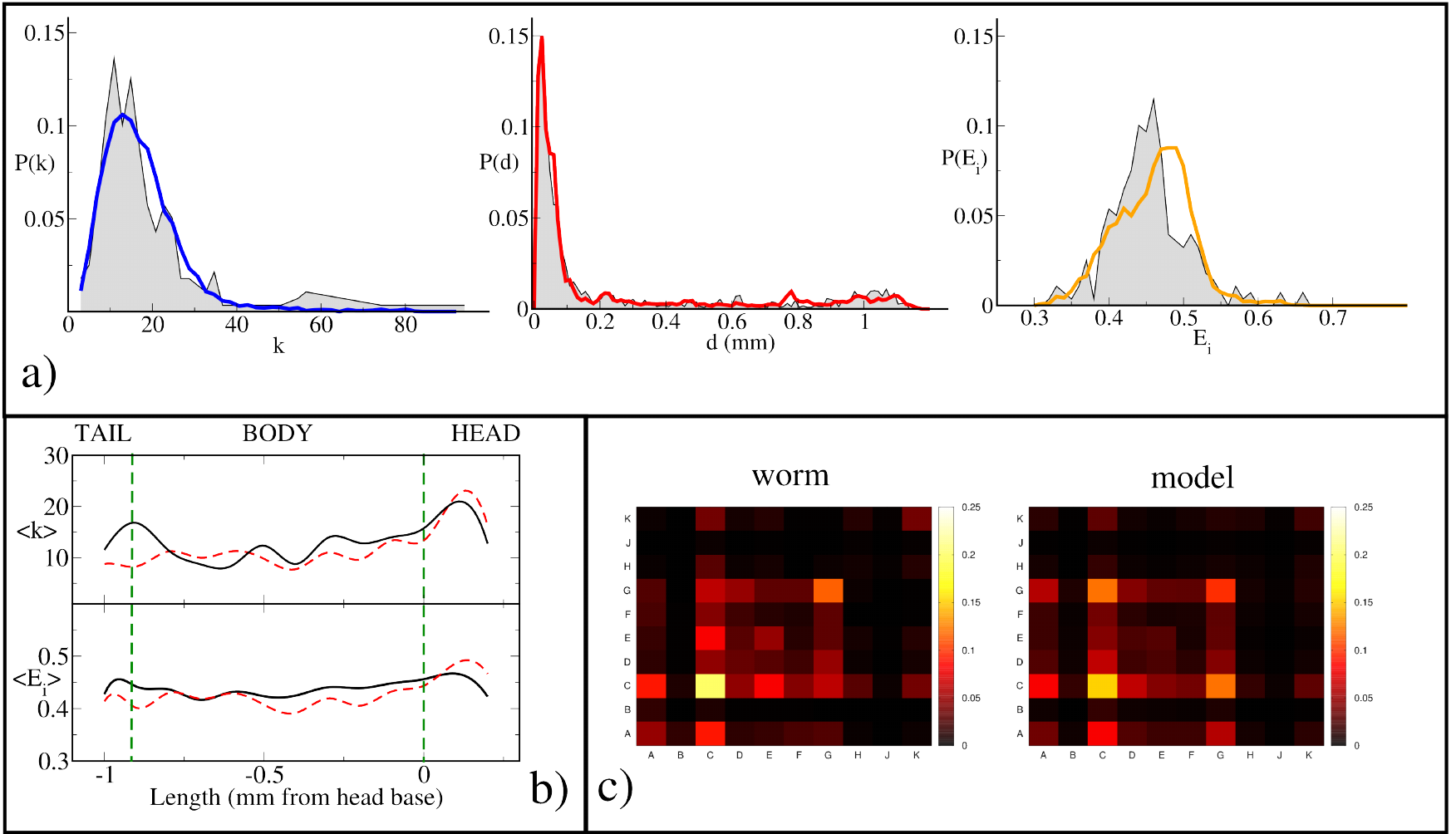}
  \end{center}
  \caption{\footnotesize\textbf{Local and mesoscopic network
      structures}. \textbf{a)}.The distributions of node degree (Left,
    blue), connection distance (Center, red) and node efficiency
    (Right, orange) of model-generated networks closely match those
    observed in the {\em C.\ elegans} neuronal network (shown in
    gray). \textbf{b)}. This panel shows how the average node degree
    (Upper) and the average node efficiency (Lower) vary along the
    length of the {\em C.\ elegans} body (solid black lines) and in
    networks generated using the ESTG model (red dashed
    lines). \textbf{c) } Networks created using the ESTG model
    (right-panel) also reproduce the pattern of intra- and
    inter-ganglia connections observed in {\em C.\ elegans}
    (Left). Brighter colors indicate higher connection density;
    letters A-K denote neuronal ganglia as defined in legend to
    Figure~\ref{fig:fig1}.}
    \label{fig:fig3}
\end{figure*}

The economical spatio-temporal growth model also provides a good
account of several other features of the adult nervous system's
organization, including the statistical distributions of node degree,
node efficiency, and edge length in the adult worm brain (see
Figure~\ref{fig:fig3}). According to the results obtained through the
computation of the Symmetrized Kullback-Leibler divergence, ESTG is
the model which most closely reproduces the distributions of node
degree, edge length and node efficiency (see Appendix Section~S5,
Table~S-III and Fig.~S3, S4 and S5).

Moreover, the model can provide a reasonable account of finer-grained
details of the adult system, such as the anatomical variation in the
average node degree and nodal efficiency along the length of the
worm. Networks simulated by the model also had a mesoscopic structure
which closely resembled the pattern of clustered connectivity between
neurons belonging to one of 10 ganglia previously defined on
biological grounds. Neurons belonging to the same ganglion in the worm
brain tend to have high connectivity with each other and relatively
sparse connectivity to neurons in other
ganglia~\cite{arenas08,arenas09}. This biological pattern and the
neurons belonging to each specific ganglion were quite accurately
reproduced by the economical spatio-temporal growth model
(Figure~\ref{fig:fig3}).
\section*{Discussion}
We have shown that a fairly simple economical model was adequate to
account for many aspects of the spatial and topological development of
the nervous system of the nematode worm, {\em C. elegans}. We describe
this generative model as economical because it represents the
formation of synaptic connections probabilistically as a trade-off
between topological value and wiring cost. More specifically, the
model accommodates the potentially competitive tendencies of each new
neuron to connect to topologically important hub neurons, which may be
a long distance away ($\sim 1$mm), versus connecting only to neurons
that are spatially adjacent ($<0.1$mm), which will conserve wiring
cost. Crucially, in estimating the connection cost between pairs of
neurons we have used prior data on the birth time of each neuron and
the progressive elongation of the worm's body to estimate the distance
between each pair of neurons at the time of synapse formation. This
measure of connection cost was traded-off against a topological bias
(preferential attachment) for new neurons to connect to high degree
hub neurons of the adult nervous system. As the worm's body
progressively elongates, the cost penalty predominates and long
distance connections, even to hub nodes, become less likely. This
simple but novel model of a dynamically evolving economical trade-off
between cost and topology has allowed us to reproduce a phase
transition in the growth of the {\em C. elegans} cellular connectome
coinciding closely with the moment of hatching, or metamorphic
transition from egg to larval stages of development. Dynamical
economical growth processes also simulated several aspects of the
configuration of the adult nervous system.

The principle that nervous systems conserve wiring cost dates back to
the seminal work of Ram\'on y Cajal in the 19th century and it has
been experimentally validated and theoretically developed extensively
since then. Many aspects of brain organization, ranging from the
placement of neurons in the adult {\em C. elegans} nervous system
\cite{chen06}, to the shape of dendritic trees \cite{cuntz10} and the
modular architecture of large-scale human brain
networks~\cite{bassett10}, have been plausibly attributed to a
parsimonious drive to minimize wiring cost. However, a strictly
cost-minimal network would have a regular, lattice-like
topology. Synaptic connections would be clustered between spatially
and topologically neighbouring neuronal nodes, with none of the long
distance axonal projections needed to mediate topologically efficient
communication between widely separated neurons. But this is not a
recognisable description of nervous system topology. In many species,
and at many scales of space and time, it has been found that brain
structural and functional networks have shorter average path length or
greater efficiency than a regular lattice.  Brain networks also
consistently have non-regular properties like high-degree hubs in a
fat-tailed degree distribution, and a modular community structure
entailing long distance inter-modular connections between neurons in
anatomically distributed modules. Many of these topological features
are more than minimally expensive or incur a premium in wiring cost;
but they can add value to the overall performance of the system. For
example, high-degree hub nodes of the {\em C. elegans} nervous system
include many of the so-called command interneurons which play a key
role in the adaptive function of coordinated forward and backward
movement of the worm \cite{varshney11,towlson13}. Topological
efficiency of human brain networks has been positively correlated with
normal variation in IQ (more intelligent people tend to have more
efficient structural and functional networks)
\cite{heuvel09,bullmore12}. Trade-offs between cost and efficiency
have been shown to be heritable properties of human brain networks
derived from functional magnetic resonance imaging (fMRI) data
\cite{fornito11}; and economical models of network formation can
reproduce the (somewhat different) statistical properties of fMRI
networks in both healthy adults and patients with schizophrenia
\cite{vertes12}. These and other observations support the general idea
that nervous systems are selected to negotiate an economical trade-off
between wiring cost (usually measured by connection distance) and
topological value (which could be measured by degree, efficiency or a
number of other network properties related to adaptive brain
function).

So the basic principles of the economical model investigated here are
not new to the neuroscience literature~\cite{bullmore12}. However,
there are several distinctive aspects of our results. Firstly, this
work is an innovative demonstration that economical models can account
for the growth of a nervous system described quite concretely and
exactly at the cellular scale of synaptic connections between neurons.
Many of the previous studies of economical trade-offs in brain
networks have been based on analysis of statistical associations
(so-called functional connectivity) between fMRI time series recorded
at different spatial locations \cite{fornito11}; or on analysis of
large-scale axonal projections rendered by tractography algorithms
applied to diffusion imaging data \cite{vandenheuvel12}. Such human
neuroimaging results indicate that economical principles may apply to
network formation at macroscopic scales, but the neuronal substrate of
networks based on imaging statistics remains unresolved. The
demonstration here of economical principles applying to a connectome
described with much greater precision at a cellular scale somewhat
validates the prior neuroimaging results. Moreover, it suggests that
the same competitive selection criteria may inform nervous system
formation over multiple spatial scales. Brain networks may have a
scale-invariant or fractal economy of organization.

More broadly, these results are innovative in demonstrating directly
how simple economical growth models can provide a reasonable account
of complex growth curves, such as the non-linear processes of nervous
system maturation and metamorphosis, from egg to adult
worm. Nematodes, like all superphylum Ecdysozoa, develop through
discrete stages (egg, several juvenile stages, adult) separated by
moulting events.  The situation in {\em C. elegans} is most closely
analogous to hemimetabolous insects (with ``incomplete
metamorphosis'') since the juvenile stages resemble the adults apart
from the absence of mating/reproductive structures.  However, each
moult can be considered metamorphic, with the L1-L2 and L4-adult
moults in particular known to involve both the addition of new cells
and formation of new synaptic links.  The special significance of the
egg-L1 transition has perhaps been less appreciated up to now, and as
such represents a unique finding of this work

Our more realistic modelling of connection cost, taking into account
the changing spatial constraints during the growth of the system, also
shines a different light on the many previous studies of connection
cost ~\cite{chen06, perez07, varshney11} in this paradigmatic complex
system.  Further work will be needed to test the hypothesis that the
specific parameters of this model correspond to discrete molecular or
genetic signals. It is imaginable, for example, that a penalty on long
distance connections could be biologically coded by the spatial
gradient of an axonally attractive molecule diffusing from neurons; or
that neurons destined to have high degree in the adult system express
distinctive cell surface markers from birth that favor synaptic
formation.

We have compared the performance of the dynamically evolving
economical model to that of a number of other models and found, as
expected theoretically, that simpler models based on preferential
attachment rules could reproduce one or other of the two phases
(quadratic or linear) of network development pre- or
post-hatching. However, only economical models which traded-off
connection distance versus preferential attachment bias could
reproduce both phases and the timing of phase transition was only
accurately reproduced by the dynamic linkage between inter-neuronal
connection distance and progressive developmental elongation of the
whole organism. For this reason, we consider that the modelling
results affirm our hypothetical prediction that development of this
cellular connectome can be accounted for by continual re-negotiation
of an economical trade-off between connection cost and the formation
of high degree hubs. This affirmation is conditional on the caveats
that not all possible models have been comparatively evaluated. It is
possible that a better model, perhaps incorporating a few more
relevant biological details (such as type of synapse, electrical or
chemical), could be developed in future.

We note that economical principles of network formation demonstrated
here for the growth of the nervous system of the nematode worm are not
necessarily limited to this system. Many other systems, besides
brains, are both spatially embedded and topologically complex. We
anticipate that economical growth models of the potentially changing
trade-offs between physical connection cost and topological value may
also contribute to future understanding of the development and
evolution of transport, computational and infrastructural systems.

\section*{Materials and Methods}
\noindent
\textbf{Data.} We have used the most up-to-date map of the {\em
  C. elegans} connectome~\cite{varshney11}, consisting of $279$
somatic neurons interconnected through $6393$ chemical synapses, $890$
gap junctions and $1410$ neuromuscular junctions. Since gap junctions
often overlap with synapses and synaptic connections are often
reciprocated, we have considered only the backbone network, where all
the synapses and gap junctions between each pair of neurons are
represented by a single undirected edge, obtaining a graph with
$N=279$ nodes and $K=2287$ edges in total (neuromuscular connections
were excluded). Information about the growth of the neuronal network,
in particular on the exact time of birth of each neuron, has been
reconstructed from recent literature~\cite{kaiser11}.

\noindent
\textbf{Linear Preferential Attachment}. The Barab\'asi and Albert
(BA) model assumes that the growth of a network is solely driven by
its topological structure, and produces random graphs with a
power--law degree distribution $p_k \sim k^{-\gamma}$, where
$\gamma\simeq 3$~\cite{barabasi99}. In the model, a new node is added
at each time and is connected to $m$  existing nodes.  The
probability for the new node $i$ to be connected to an existing node
$j$ is a linear function of the degree $k_j$, namely:
\begin{equation}
  \Pi^{BA}_{i\rightarrow j}= k_j/2K
\end{equation}
where $K$ denotes the total number of links when the new node arrives.
Since each node chooses $m$ neighbors to connect with, the total
number of links increases linearly with the size of the network, and
the average node degree remains constant.

\noindent
\textbf{Accelerated Topological Growth}. Traditionally, network growth
is said to be accelerated if the average node degree
increases with the size of the network. Acceleration has been observed
in many complex networks and different models of scale-free networks
with acceleration have been proposed so far~\cite{mendes01}. We have
considered two different accelerated growth models. In the first
model, called Binomial Accelerated Growth (BAG), a new node $i$ tries
to establish a connection with each of the existing nodes, and a link
to node $j$ is created with probability $p$, namely:
\begin{equation}
  \Pi^{BAG}_{i\rightarrow j}= p
\end{equation}
The BAG model produces networks in which the number of links increases
as the square of $N$. In fact, the expected number of links
established when the network has $N$ nodes is:
\begin{equation}
  K(N) = p\sum_{i=1}^{N}(i-1) = p\frac{N(N-1)}{2}
\end{equation}
The BAG model produces networks with a binomial degree distribution,
since it is equivalent to an Erd\H{o}s-R\'enyi random graph model,
where each of the $N(N-1)/2$ potential links appears with probability
$p$~\cite{erd60}.

We introduced also a second model of accelerated growth, called
Hidden-variable Accelerated Growth (HAG). In general, hidden-variable
models produce networks with a prescribed degree distribution: the HAG
model grows random networks having -- on average -- the same degree
distribution observed in the adult {\em C.\ elegans} neural
network. The model works as follows. We assign to each node $j$ of the
network, once and for all, a hidden variable $h_j$. In
particular, we set $h_j={k_j}^{\text{ad}}$, where ${k_j}^{\text{ad}}$
is the degree of node $j$ in the adult worm. When a new node $i$
arrives, it tries to establish a link with each of the nodes in the
network, and a link to node $j$ is created with probability:
\begin{equation}
  \Pi^{HAG}_{i \rightarrow j} =p\frac{h_j}{h_{max}}
\end{equation}
where $h_{max}$ is the maximum of $h_j$ over $j$, and $p$ is
appropriately selected in order to reproduce the final number of
links. It is possible to prove that the final degree of a node $i$
over different network realizations is Poisson distributed around an
average value equal to ${k_i}^{\text{ad}}$. Consequently, networks
produced by HAG show an accelerated growth similar to that generated
by the BAG model, while also preserving the actual degree distribution
of the {\em C.\ elegans} neural network.

\noindent
\textbf{ESG}.  To create networks embedded in Euclidean
space~\cite{barthelemy03,barthelemy11}, we considered the economical
spatial growth model, which is based on a trade-off between the
tendency to create topologically important connections to hubs and the
physical distance between neurons. When a new node $i$ arrives, it is
placed in the position it occupies in the adult worm, and a link to
each of the existing nodes is created with probability:
\begin{equation}
  \Pi^{ESG}_{i \rightarrow j} = \frac{h_{j}}{h_{max}}
  e^{-\frac{d_{ij}^{\text{ad}}}{\delta}}
\end{equation}
where the values $h_j$ are assigned as in the HAG model and $\delta$
is a parameter tuning the typical connection distance.  Namely, the
probability of creating a link exponentially decreases with the
Euclidean distance $d_{ij}^{\text{ad}}$ that separates $i$ and $j$ in
the adult worm, and is weighted by the hidden variable
$h_j={k_j}^{\text{ad}}$ (in order to preserve the actual degree
distribution of the {\em C.\ elegans} neural network).

\noindent
\textbf{ESTG}.  The Economical Spatio-Temporal Growth model, using
information about the length of the worm at different stages, takes
into account the actual spatial position of each neuron, while the
worm grows over time. When a new node $i$ arrives, it is placed in the
position it occupies in the {\em C.\ elegans} neural network at time
$t$, and a link to each of the existing nodes is created with
probability:
\begin{equation}
  \Pi^{ESTG}_{i \rightarrow j} =
  \frac{h_{j}}{h_{max}}e^{-\frac{d_{ij}(t)}{\delta}}
\end{equation}
where the values $h_j$ are assigned as in the HAG model, and $\delta$
is a parameter tuning the typical edge length.  Notice that the
probability to establish a link depends on the time at which node $i$
appears, since the distance $d_{ij}(t)$ depends on the relative
positions of $i$ and $j$, which change over time due to elongation of
the worm's body.  We considered the real length of the worm at each
time, and we estimated the position of each node at that time using
linear interpolation and assuming a uniform expansion of the worm
along the longitudinal axis.

\noindent
\textbf{Parameter Tuning}. The first requirement of any suitable model
for the {\em C. elegans} neuronal network growth is to produce
networks having $N=279$ nodes and, on average, $K=2287$ edges, as
observed in the adult worm.  We used Monte Carlo simulations and
iterative bisection to identify the interval in the parameter space
for which the expected total number of edges $\tilde{K}$ of the
generated networks was equal to $2287\pm 1\%$; see Appendix Section
S3 for methodological details and the optimal parameter values for
each of the eight models in  Appendix Table~S-I.

\noindent
\textbf{Degree Distribution}. Given an undirected graph $G(V,E)$
associated with the symmetric adjacency matrix $A=\{a_{ij}\}$, the
degree of a node $i$ is defined as the number of edges incident on
$i$, and is denoted by $k_i=\sum_{j}a_{ij}$. The degree distribution
$P(k)$ of the graph indicates, for each value of $k$, the probability
of finding a node whose degree is equal to $k$.

\noindent
\textbf{Connection Distance Distribution}. Given two directly connected nodes
$i$ and $j$ of a spatially-embedded network, we define the distance of
the edge $(i,j)$ as the Euclidean distance $d_{ij}$ separating node
$i$ and node $j$. The distance distribution $P(d)$ is
the probability of finding an edge whose distance is exactly equal to
$d$.

\noindent
\textbf{Node and Graph Efficiency}. Given an undirected and unweighted
graph $G$, the efficiency of a node is defined as:
\begin{equation}
  E_{i} = \frac{1}{N-1}\sum_{\stackrel{j=1}{j\neq
      i}}^{N}\frac{1}{\lambda_{ij}}
\end{equation}
where $\lambda_{ij}$ is the path length between node $i$ and node $j$,
measured as the number of edges in the shortest path connecting $i$ to
$j$. The smaller $\lambda_{ij}$, the larger the contribution of node
$j$ to the efficiency of $i$.  The efficiency of a graph is defined as
the average efficiency of its nodes.

\section*{acknowledgments}
V.N. and V.L. acknowledge the support of the EU Project LASAGNE,
Contract no. 318132 (STREP). Behavioural \& Clinical Neuroscience
Institute is supported by the Medical Research Council (UK) and the
Wellcome Trust.

\newpage

\section*{Appendix}

\renewcommand\theequation{{S-\arabic{equation}}}
\renewcommand\thetable{{S-\Roman{table}}}
\renewcommand\thefigure{{S-\arabic{figure}}}
\renewcommand\thesection{{S\arabic{section}}}
\setcounter{figure}{0}
\setcounter{section}{0}
\setcounter{equation}{0}

\noindent
\textbf{Section S1. Location of neurons born after hatching}
\addcontentsline{toc}{section}{S1. Location of neurons born after hatching}

\begin{figure*}[!ht]
  \begin{center}
    \includegraphics[width=7in]{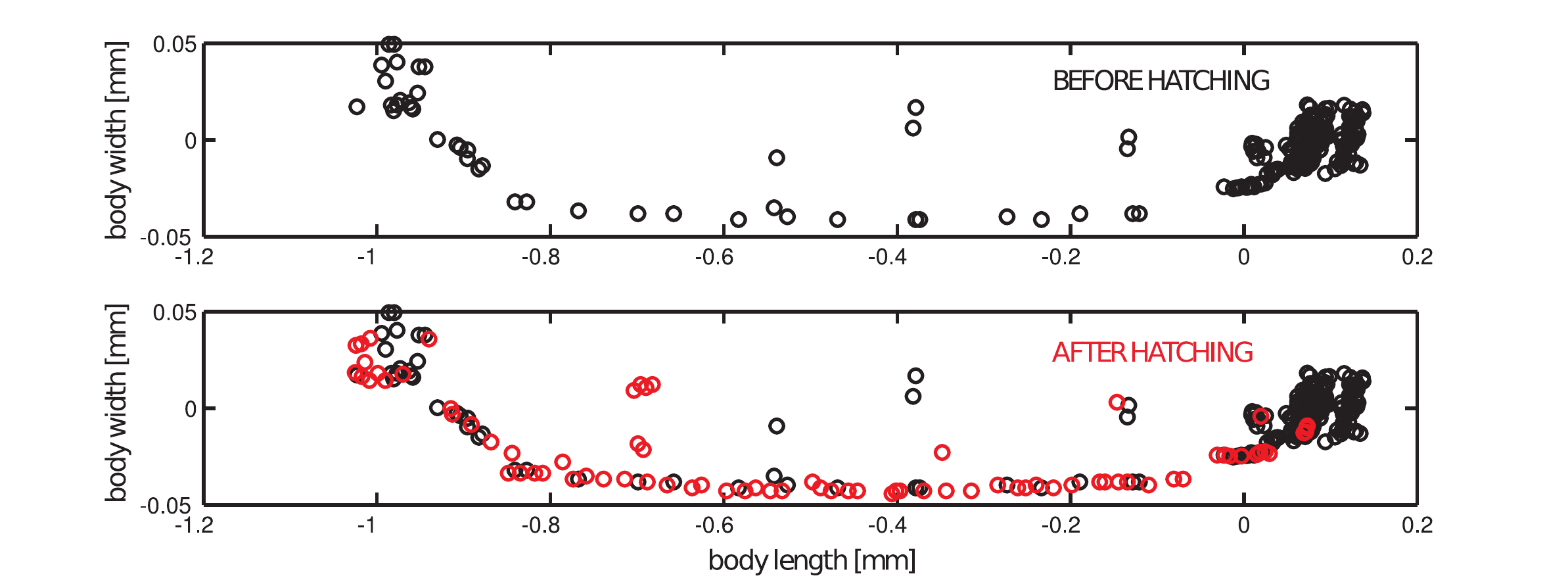}
  \end{center}
  \caption[Position of neurons born after hatching]{\textbf{Position
      of neurons born after hatching.} The large majority of neurons
    born after hatching are located throughout the worm's body, while
    most of the neurons born before hatching are concentrated in the
    head and in the tail. The x-axis represents the distance in
    millimeters from the base of the head. Positive values indicate
    points in the worm's head, while negative values correspond to the
    body and the tail.}
  \label{fig:s_fig1}
\end{figure*}

In Fig.~\ref{fig:s_fig1} we show the spatial configuration of neurons
before and after hatching. Notice that the majority of the neurons
born before hatching are concentrated in the head and in the tail
region, while most of the neurons appearing after hatching are instead
placed in the body to form the ventral cord. This explains the
relative higher distance from newly added neurons to existing ones
observed after hatching.

\bigskip
\noindent
\textbf{Section S2. Additional one-parameter models}
\addcontentsline{toc}{section}{S2. Additional one-parameter models}

We present here three additional growth models which have been tested
during this study, namely the Simple Spatial Growth (SSG), Spatial
Growth with Elongation (SGE) and Power-law Economical Growth (PEG). We
also discuss their ability to reproduce the developmental growth of
the {\em C. elegans} neuronal network, and we will compare them with
the other five models described in the main text,
i.e. Barab\'asi-Albert (BA), Binomial Accelerated Growth (BAG),
Hidden-variable Accelerated Growth (HAG), Economical Spatial Growth
(ESG) and Economical Spatio-Temporal Growth (ESTG). Notice that all
the models considered in this study have only one free
parameter. Nevertheless some of these models, and in particular the
ESTG, are exceptionally accurate at reproducing the structure and
development of the {\em C. elegans} neuronal network.

\textit{Simple Spatial Growth (SSG).} This model makes the assumption
that upon arrival a new node $i$ is placed in the same position at
which it appears in the adult worm. Then, node $i$ creates an edge to
each of the already existing nodes $j$ with probability:
\begin{equation}
    \Pi^{SP}_{i\rightarrow j}= e^{-\frac{d_{ij}^{\text{ad}}}{\delta}}
\end{equation}
where $d_{ij}^{ad}$ is the distance between node $i$ and node $j$ in
the adult worm and $\delta$ is a parameter tuning the typical edge
length. Since the connection probability decreases exponentially with
the distance between nodes in the adult worm, the resulting networks
exhibit very few  medium- and high-distance
links, which are instead relatively frequent in the real {\em C. elegans}
neuronal networks.

\textit{Spatial Growth with Elongation (SGE).} This model uses
information about the length of the worm at different stages. The node
$i$ arriving in the network at time $t$ is placed in the position it
occupies in the neural network at that time, and the probability for
$i$ to connect to an existing node $j$ is defined as:
\begin{equation}
  \Pi^{SPE}_{i\rightarrow j} = e^{-\frac{d_{ij}(t)}{\delta}}
\end{equation}
where $d_{ij}(t)$ is the distance between node $i$ and node $j$ at
time $t$ and $\delta$ is a parameter. Notice that $d_{ij}(t)$ is a
function of time, so that the probability to create an edge between a
newly arrived node $i$ and an existing node $j$ depends on the time at
which node $i$ arrives in the network and on the relative positions of
$i$ and $j$ at that time. This makes possible the creation of edges
between nodes which are actually separated by a relatively large
distance in the adult worm but have been closer in space in earlier
developmental stages.

\textit{Power-law Economical Growth (PEG).} This model implements a
trade-off between the tendency to create edges to hubs and the
relative distance of the nodes, and takes into account the elongation
of the worm during development. Differently from the Economical
Spatio-Temporal Growth model presented in the main text, in which the
connection probability is a decreasing exponential function of
distance, in this model the probability to connect to a distant node
decreases as a power-law:
\begin{equation}
  \Pi^{PEG}_{i\rightarrow j} = \frac{h_{j}}{h_{max}}\left[1 -
    \left(\frac{d_{ij}(t)}{L_t}\right)^{\alpha}\right]
\end{equation}
Here, $h_{j}$ is the hidden degree of node $j$, which is set equal to
the degree of node $j$ observed in the adult worm, while $h_{max}$ is
maximum node degree in the adult neural network. As for the ESTG
model, $d_{ij}(t)$ is the distance between node $i$ and node $j$ in
the worm at time $t$.  $L_t$ is the total worm length at time $t$ and
$\alpha$ is the exponent of the power-law. Notice that the attachment
probability $\Pi^{PEG}$ approaches $0$ when the distance $d_{ij}(t)$
is comparable with the length of the worm, while the hidden degree of
the destination node plays a more important role if the two nodes are
closer in space. Thanks to the preferential attachment term, based on
the hidden degree of the nodes, this model tends to preserve the
degree distribution of the original network.

\bigskip
\noindent
\textbf{Section S3. Parameter tuning}
\addcontentsline{toc}{section}{S3. Parameter tuning}

In this study we considered only one-parameter randomized growth
models. In general, a randomized model generates an ensemble of graphs
having certain characteristics. If the model has a tunable parameter,
each value of the parameter generates a family of graphs sharing
similar structural properties. For instance, the Binomial Accelerated
Growth (BAG) model produces networks in which the number of edges
grows quadratically with the number of nodes, but the expected number
of edges in the final network, i.e. when the number of nodes is equal
to $N=279$, depends on the actual value of the attachment probability
$p$.

Since a randomized one-parameter model generates a family of graphs
for each value of the parameter, its ability to reproduce the
structure of a given network cannot be assessed through a direct
comparison of the original graph with a single realization of the
model. Instead, the comparison should be performed by taking into
account the expected structural properties of the ensemble of networks
generated by the model, for each value of the parameter, averaging
over a sufficiently large number of realizations.  The first
requirement of any suitable model for the {\em C. elegans} neural network
growth is to produce networks having $N=279$ nodes and, on average,
$K=2287$ edges. This constraint has been used to find the optimal
parameter of each considered model.

We employed a two-step parameter optimization process. In the first
step we used a Monte-Carlo approach to identify the interval in the
parameter space for which the expected total number of edges
$\tilde{K}$ of the generated networks was equal to $2287\pm 5\%$. In
this step, we considered $20$ networks for each value of the
parameter. In the second step we iteratively shrunk the parameter
interval using the bisection method, in order to identify the value
for which the difference between $\tilde{K}$ and $K=2287$ was smaller
than $1\%$. In this step we generated $500$ networks for each value of
the parameter. The optimal parameter values for each of the eight
models are reported in Table~\ref{tab:table1}.
\begin{table}[!h]
\begin{tabular}{|l|l|}
  \hline
  \textbf{Model} & \textbf{Optimal parameter} \\ \hline
    BAG & $p=0.0575$\\
    HAG & $p=0.302$\\
    BA & $m_0=8, m=8$\\
    SSG & $\delta=0.01365$\\
    SGE & $\delta=0.00235$\\
    PEG & $\alpha=0.0232$\\
    ESG & $\delta=0.0858$\\
    ESTG & $\delta=0.0126$\\ \hline
\end{tabular}
\caption[Optimal model parameters]{\textbf{Optimal model parameters.}
  The optimal parameter of a model guarantees the generation of
  networks having the same number of edges as the {\em C. elegans}
  adult neural network, with an error smaller than $1\%$.}
    \label{tab:table1}
\end{table}

\bigskip
\noindent
\textbf{Section S4. Model comparison}
\addcontentsline{toc}{section}{S4. Model comparison}

Since our aim was to reproduce as closely as possible the
developmental growth of the {\em C. elegans} neuronal network, and in
particular the abrupt transition in the number of edges in the graph
as a function of the number of nodes, we defined a measure to quantify
how closely each model matches the curve $\mathcal{K}(N)$, which
indicates the number of edges in the {\em C. elegans} neuronal network when $N$
nodes have been born.

We denote by $\mathcal{K}_{M}(N)$ the family of curves of $K$ over $N$
obtained using a certain model $M$ and setting the value of the model
parameter according to Table~\ref{tab:table1}.  We computed, for each
value of $N$, the expected number $\mu(\mathcal{K}_M(N))$ of edges in
the network generated by model $M$ when $N$ nodes have been added to
the graph, averaging over $500$ realizations. Using this notation,
$\mu(\mathcal{K}_{M}(100))$ is the expected number of edges in the
graphs generated by model $M$ when the first $N=100$ nodes have been
added to the graph.

\begin{figure*}[p]
  \begin{center}
    \includegraphics[width=6in]{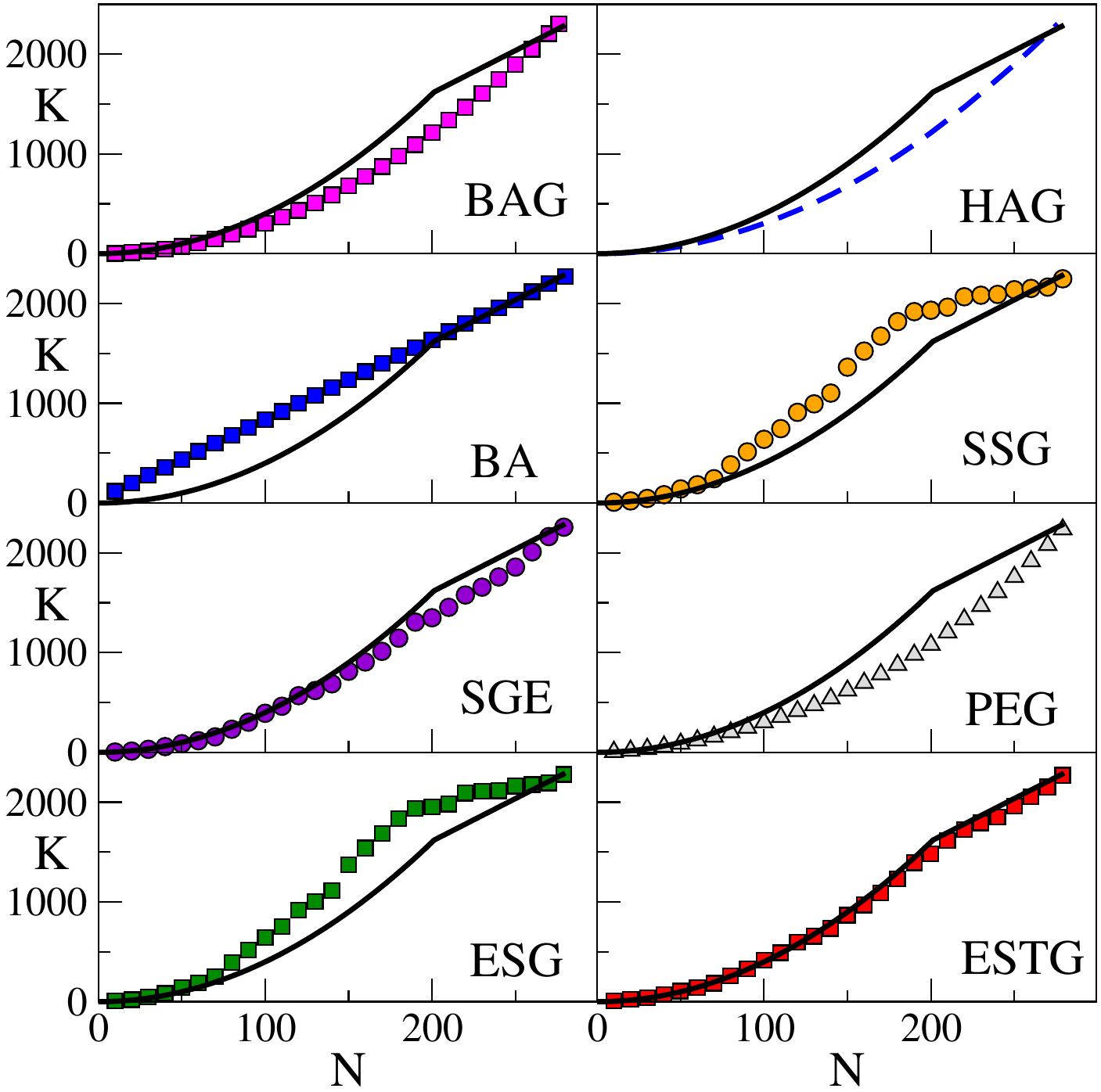}
  \end{center}
  \caption[Growth curves]{\textbf{Growth curves.} The average total
    number of edges $\mathcal{K}_{M}(N)$ as a function of $N$ for each
    of the eight models. The original growth curve of the {\em
      C. elegans} neural network is reported for reference in each
    panel, as a solid black line. The SSG, SGE, ESG and ESTG models
    exhibit a transition from a quadratic to a linear increasing
    regime, but only ETSG is able to closely match the growth curve
    observed in the original graph.}
  \label{fig:s_fig2}
\end{figure*}

In Fig.~\ref{fig:s_fig2} we report the average curve
$\mu(\mathcal{K}_M(N))$ for each of the eight considered models,
together with the original curve $\mathcal{K}(N)$ corresponding to the
growth of the {\em C. elegans} neural network. By visual inspection, we
conclude that the model which best fits the developmental growth of
the original network and the phase transition at hatching is the
Economical Spatio-Temporal Growth.  In order to quantify the
discrepancy between $\mathcal{K}(N)$ and $\mathcal{K}_M(N)$ we
computed, for each model and for each value of $N$, the difference
$\xi(N)$:
\begin{equation}
  \xi(N) = \left|\mathcal{K}(N) - \mu(\mathcal{K}_M(N))\right|
\end{equation}
and we considered the expected value $\mu[\xi(N)]$ and the standard
deviation $\sigma[\xi(N)]$ of $\xi(N)$. In Table~\ref{tab:table2} we
report the values of $\mu[\xi(N)]$ and $\sigma[\xi(N)]$ for the eight
models considered. In general, smaller values of $\mu[\xi(N)]$ and
$\sigma[\xi(N)]$ indicate a closer match of the original growth curve.
In agreement with the conclusions drawn after visual inspection of
Fig~\ref{fig:s_fig2}, which suggested that ESTG was the model which
most closely reproduced the growth curve, the smallest values of
$\mu[\xi(N)]$ and $\sigma[\xi(N)]$ are indeed obtained by the
Economical Spatio-Temporal Growth model. The networks generated by all
the other models fail to follow the original growth curve by a large
extent, and they consequently exhibit larger values of $\mu[\xi(N)]$
and $\sigma[\xi(N)]$.
\begin{table}[!h]
  \begin{tabular}{|l|r|r|}
    \hline
    \textbf{Model} & \textbf{$\mu[\xi(N)]$} & \textbf{$\sigma[\xi(N)]$} \\ \hline 
    BAG & 154.2 & 123.7\\
    HAG & 154.2 & 123.7\\
    BA & 216.7 & 150.7\\
    SSG & 205.2 & 167.1\\
    SGE & 89.5 & 73.7\\
    PEG & 209.4 & 168.4\\
    ESG & 215.6 & 172.9\\
    \rowcolor{green}ESTG & \textbf{37.3} & \textbf{31.6}\\
    \hline
  \end{tabular}
  \caption[Quality of growth fit]{\textbf{Quality of growth
      fit.}Average and standard deviation of the point-to-point
    difference between the observed growth curve $\mathcal{K}(N)$ and
    the average curve corresponding to each of the eight considered
    models. The model parameters are set according to
    Table~\ref{tab:table1}. The smaller the value of $\mu[\xi(N)]$,
    the more closely a model can reproduce the growth of the {\em
      C. elegans} neural network. The Barabasi-Albert model (BA)
    exhibits the highest average point-to-point distance, while the
    Economical Spatio-Temporal Growth model (ESTG) largely outperforms
    all the other models.}
  \label{tab:table2}
\end{table}

\begin{table*}[!ht]
  \begin{tabular}{|l|r|r|r|}
    \hline 
    \textbf{Model} & $D_{KL}\left(P(k), P_{M}(k)\right)$ &
    $D_{KL}\left(P(d), P_M(d)\right)$ & $D_{KL}\left(P(E_i),
    P_M(E_i)\right)$ \\ \hline
    BAG & \cellcolor{red}0.875 & 0.346 & \cellcolor{orange}0.966\\
    HAG & 0.301 & 0.290 & 0.545\\
    BA & 0.309 & \cellcolor{yellow} 0.176 & 0.226\\
    SSG & \cellcolor{orange}0.710 & \cellcolor{red}0.884 & 0.611\\
    SGE & 0.428 & 0.269 & \cellcolor{red}1.447\\
    PEG & \cellcolor{yellow}0.149 & 0.322 & \cellcolor{green}0.214\\
    ESG & 0.708 & \cellcolor{orange}0.685 & 0.361\\
    ESTG & \cellcolor{green}0.143 & \cellcolor{green}0.099 & \cellcolor{yellow}0.223\\
    \hline
  \end{tabular}
  \caption[Kullback-Leibler divergence]{\textbf{Kullback-Leibler
      divergence.} The symmetrized Kullback-Leibler divergence between
    the degree, edge length and node efficiency distributions of the
    adult {\em C. elegans} neural network and the corresponding
    average distributions of the networks generated through each of
    the eight models. Smaller values of symmetrized divergence
    indicate higher similarity between the two distributions.  The
    best and second-best values are highlighted in green and yellow,
    respectively, while the worst and second-worst are marked in red
    and orange, respectively. BAG and SSG exhibit the worst values of
    divergence. Interestingly, besides being the best model at fitting
    the developmental growth of the {\em C. elegans} neural network
    (as shown in Fig.~\ref{fig:s_fig2} and in Table~\ref{tab:table2})
    ESTG performs more consistently than any of the other models in
    reproducing the structural properties of the adult worm's nervous
    system.}
  \label{tab:table3}
\end{table*}

\bigskip
\noindent
\textbf{Section S5. Node degree, edge length and node efficiency}
\addcontentsline{toc}{section}{S5. Node degree, edge length and node
  efficiency}

Here we compare the structure of the networks produced by each of the
eight models described in this study with that observed in the adult
{\em C. elegans} neural network, by using three classical network
metrics. The first metric is the degree distribution. Given an
undirected graph $G(V,E)$ associated with the symmetric adjacency
matrix $A=\{a_{ij}\}$, the degree of a node $i$ is defined as the
number of edges incident on $i$, and is denoted by
$k_i=\sum_{j}a_{ij}$. The degree distribution $P(k)$ of the graph
indicates, for each value of $k$, the probability of finding a node
whose degree is equal to $k$. The second metric is the distribution of
connection distances. Given two directly connected nodes $i$ and $j$ of a
spatially-embedded network, we define the distance of the edge $(i,j)$
as the Euclidean distance $d_{ij}$ separating node $i$ and node
$j$. The associated distance distribution $P(d)$ is the probability
of finding an edge whose distance is exactly equal to $d$. The third
metric is node efficiency. Given an undirected and unweighted graph
$G$, the efficiency of a node is defined as:
\begin{equation}
  E_{i} = \frac{1}{N-1}\sum_{\stackrel{j=1}{j\neq
      i}}^{N}\frac{1}{\lambda_{ij}}
\end{equation}
where $\lambda_{ij}$ is the distance between node $i$ and node $j$,
measured as the number of edges in the shortest path connecting $i$ to
$j$. The node efficiency of $i$ measures how easy it is to reach any
other node in the graph by starting from $i$ and traveling across
shortest paths. In general, the smaller the distance between $i$ and
$j$, the higher the contribution of $j$ to the efficiency of node $i$.
If the graph is not connected and node $i$ and $j$ belong to two
different connected components then there exists no path connecting
them. In this case, the distance $\lambda_{ij}$ is conventionally set
to $\infty$, and the contribution of node $j$ to the efficiency of $i$
is equal to $1/\infty\equiv 0$.

\begin{figure*}[p]
  \begin{center}
    \includegraphics[width=6in]{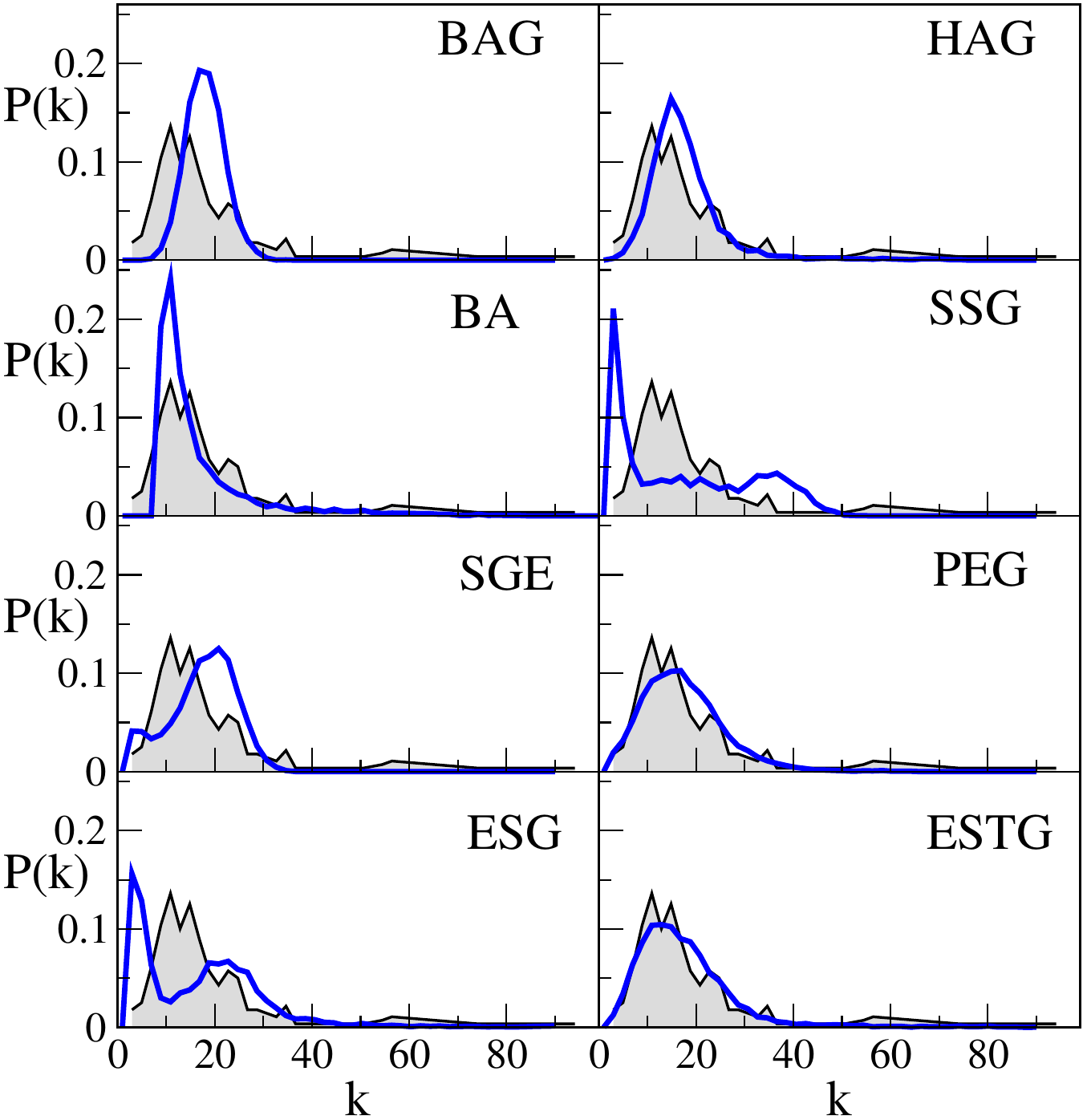}
  \end{center}
  \caption[Degree distributiuons]{\textbf{Degree distributions.} The average degree
    distribution of the networks generated by each of the eight
    models. The degree distribution of the adult {\em C. elegans} neural
    networks is reported in each panel in shaded gray, for
    comparison. Only the models based on hidden-variables, i.e. HAG,
    PEG and ETSG, are able to reproduce the degree distribution of the
    worm more closely. In the BA, SSG and ESG models low-degree nodes
    are over-represented, while in the BAG and SGE models low-degree nodes are
    substantially under-represented.}
  \label{fig:s_fig3}
\end{figure*}

\begin{figure*}[p]
  \begin{center}
    \includegraphics[width=6in]{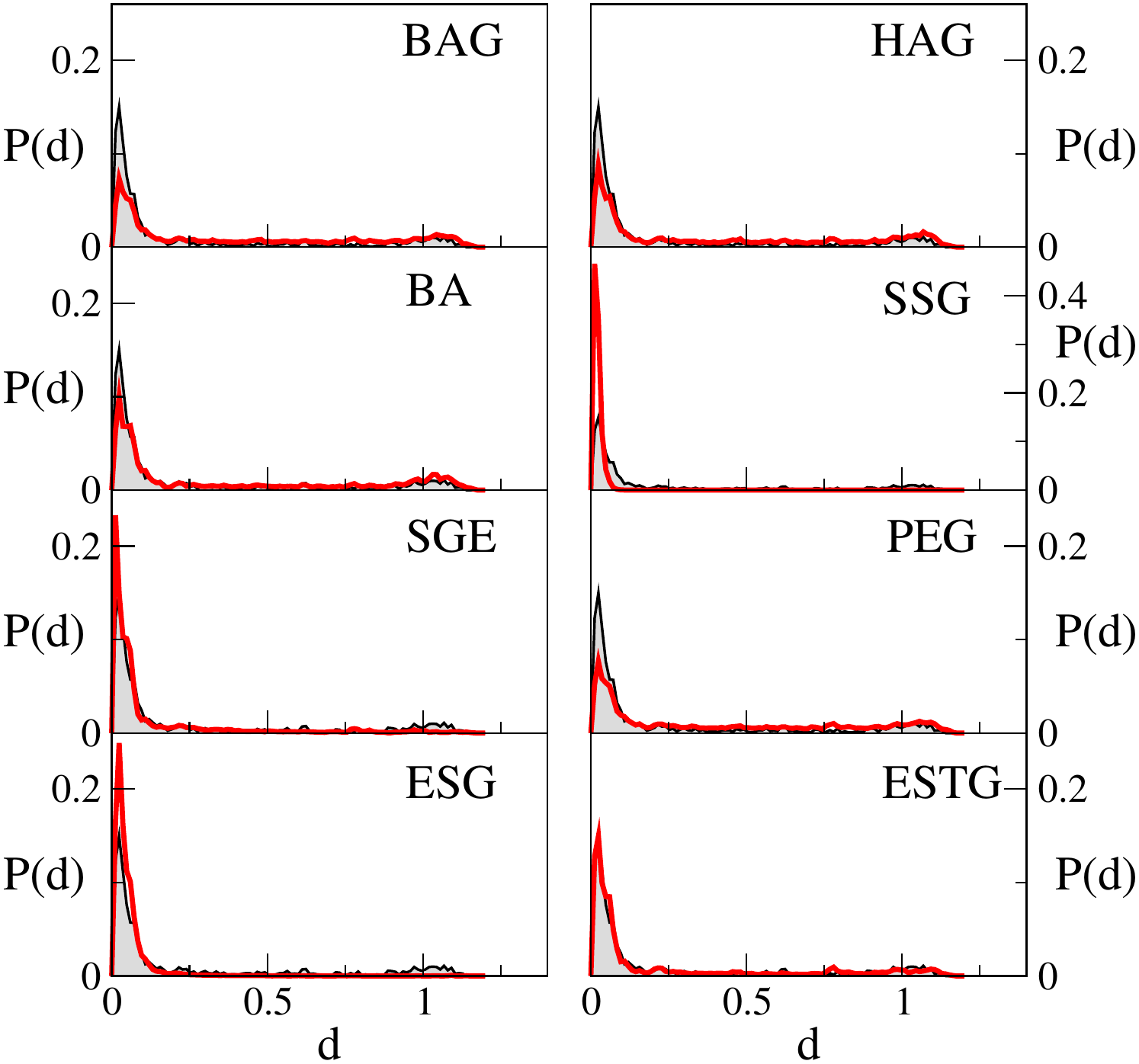}
  \end{center}
  \caption[Edge length distribution]{\textbf{Edge length
      distributions.} The average distribution of edge length in the
    networks generated by each of the eight models, compared with the
    distribution of edge length observed in the adult {\em C. elegans}
    network (reported in shaded gray). BAG, HAG, BA and PEG produce
    networks with substantially longer links, while SSG, SGE and ESG
    exhibit a substantially larger percentage of short links (notice
    the different scale of the y-axis in the SSG panel). The only
    model which closely matches the distribution of edge length is
    ESTG.}
  \label{fig:s_fig4}
\end{figure*}

\begin{figure*}[p]
  \begin{center}
    \includegraphics[width=6in]{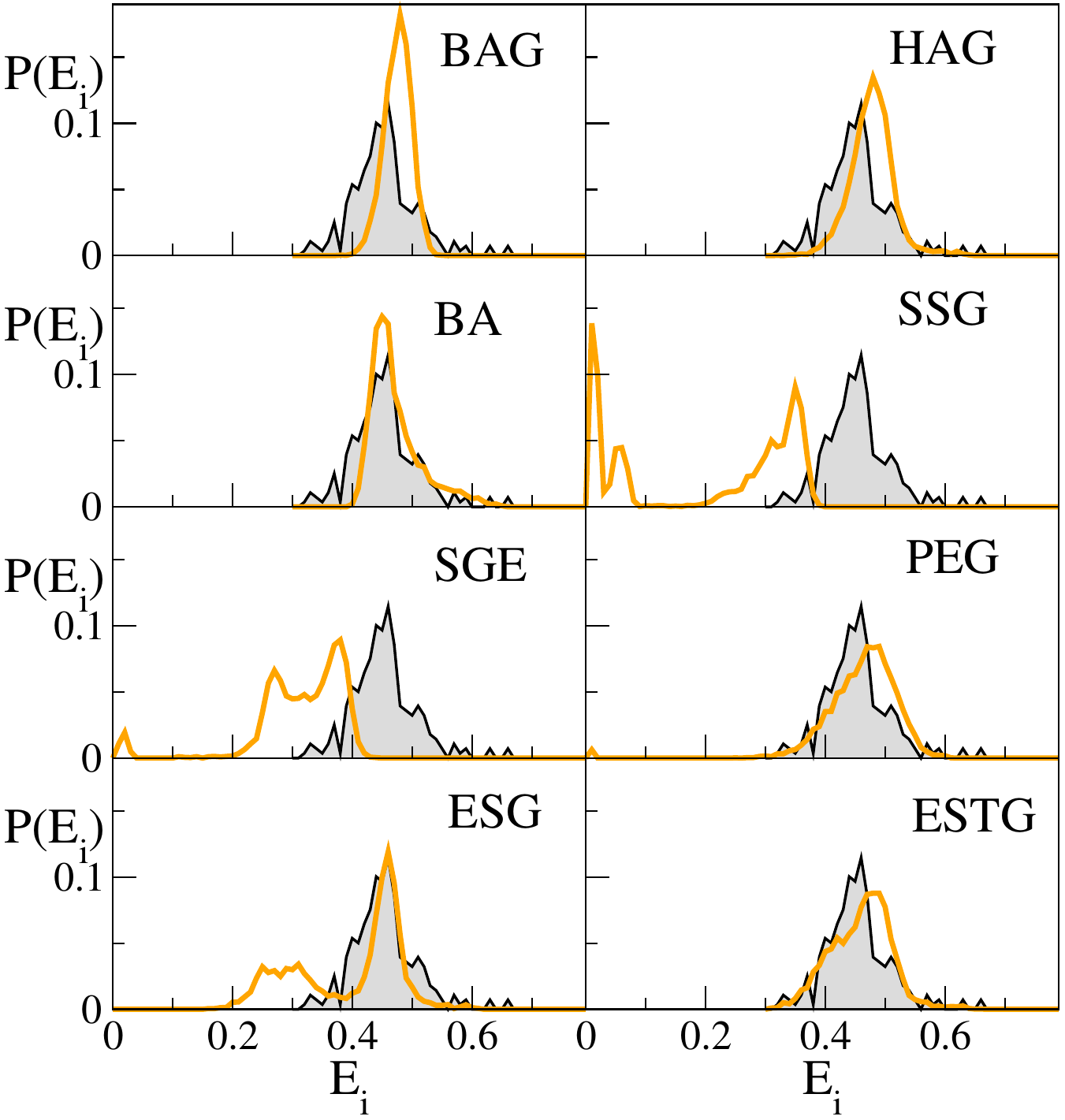}
  \end{center}
  \caption[Node efficiency distribution]{\textbf{Node efficiency
      distribution.} The distribution of node efficiency of networks
    generated with each of the eight models, compared with that
    observed in the adult {\em C. elegans} (shaded gray). Both BAG and
    HAG produce binomial distributions of edge efficiency; for SSG,
    SGE and ESG models the distribution of efficiency is skewed
    towards smaller values while BA is able to capture the peak around
    $E_{i}=0.47$. PEG and ESTG reproduce the original distribution in
    a more balanced way, even if nodes with efficiency around $0.5$
    are substantially over-represented while the peak around $0.47$ is
    missing.}
  \label{fig:s_fig5}
\end{figure*}

In Fig.~\ref{fig:s_fig3},~\ref{fig:s_fig4} and~\ref{fig:s_fig5} we show,
respectively, the average degree distribution, length distribution and
node efficiency distribution of the networks generated by each of the
eight models, together with those observed in the adult {\em C. elegans}
neural network (reported in each panel in shaded grey). By visual
inspection, we notice that ESTG seems to be the model which most
closely reproduces all these distributions.

In order to quantify the difference between the distributions of
degree, length and node efficiency of synthetic graphs with those of
the {\em C. elegans} neural network we used the Kullback-Leibler
divergence. Given two probability distributions $P=\{p_i\}$ and
$Q=\{q_i\}$, the Kullback-Leibler divergence of $Q$ from $P$ is
defined as:
\begin{equation}
  D_{KL}(P||Q) = \sum_{i}p_i \log{\frac{p_i}{q_i}}
\end{equation}
The Kullback-Leibler divergence measures the information lost when $Q$
is used as an approximation of $P$, and is non-symmetric,
i.e. $D_{KL}(P||Q)\neq D_{KL}(Q||P)$. Since we are interested in
measuring the similarity between two distributions, and not the
relative information lost when using one of them as a predictor of the
other, we opted for the symmetrized Kullback-Leibler divergence, which
is defined as follows:

\begin{equation}
  D_{KL}(P,Q) = \frac{D_{KL}(P||Q) + D_{KL}(Q||P)}{2}
\end{equation}

In general, the smaller the value of $D_{KL}(P,Q)$, the more similar
the two distributions $P$ and $Q$. If we denote by $P(c)$ the
distribution of the generic quantity $c$ in the {\em C. elegans} neural
network and by $P_{M}(c)$ the distribution of the same quantity $c$ in
networks generated through model $M$, the symmetrized Kullback-Leibler
divergence between $P(c)$ and $P_M(c)$ is denoted as
$D_{KL}\left(P(c), P_M(c)\right)$. In Table~\ref{tab:table3} we
report, for each model, the values of the symmetrized Kullback-Leibler
divergence between the degree, edge length and node efficiency
distributions of the adult {\em C. elegans} neural network and the networks
generated by each of the eight models, which are respectively denoted
by $D_{KL}\left(P(k), P_M(k)\right)$, $D_{KL}\left(P(d),
P_M(d)\right)$ and $D_{KL}\left(P(E_i), P_M(E_i)\right)$.  The best
and the second-best value of $D_{KL}(P,Q)$ for each metric are
highlighted in green and yellow, respectively. Notice that the
smallest values of the symmetrized Kullback-Leibler divergence are
consistently obtained by the ESTG model, with the only exception being
node efficiency for which PEG outperforms ESTG by a small amount.

\begin{thebibliography}{10}
\expandafter\ifx\csname url\endcsname\relax
  \def\url#1{\texttt{#1}}\fi
\expandafter\ifx\csname urlprefix\endcsname\relax\def\urlprefix{URL }\fi
\providecommand{\bibinfo}[2]{#2}
\providecommand{\eprint}[2][]{\url{#2}}


\bibitem{albert02} \bibinfo{author}{Albert, R.} \&
  \bibinfo{author}{Barab\'asi, A.-L.}  \newblock
  \bibinfo{title}{Statistical mechanics of complex networks}.
  \newblock \emph{\bibinfo{journal}{Rev. of Mod. Phys.}}
  \textbf{\bibinfo{volume}{74} (1)}, \bibinfo{pages}{47--97}
  (\bibinfo{year}{2002}).

\bibitem{boccaletti06} \bibinfo{author}{Boccaletti, S.},
  \bibinfo{author}{Latora, V.}, \bibinfo{author}{Moreno, Y.},
  \bibinfo{author}{Chavez, M.} \& \bibinfo{author}{Hwang, D.-U.}
  \newblock \bibinfo{title}{Complex networks: Structure and dynamics}.
  \newblock \emph{\bibinfo{journal}{Phys. Rep.}}
  \textbf{\bibinfo{volume}{424}}, \bibinfo{pages}{175--308}
  (\bibinfo{year}{2006}).

\bibitem{bullmore09}
\bibinfo{author}{Bullmore, E.} \& \bibinfo{author}{Sporns, O.}
\newblock \bibinfo{title}{Complex brain networks: graph theoretical analysis of
  structural and functional systems}.
\newblock \emph{\bibinfo{journal}{Nat Rev Neurosci}}
  \textbf{\bibinfo{volume}{10}}, \bibinfo{pages}{186--198}
  (\bibinfo{year}{2009}).

\bibitem{barabasi12}
\bibinfo{author}{Barab\'asi, A.-L.}
\newblock \bibinfo{title}{The network takeover}.
\newblock \emph{\bibinfo{journal}{Nature Physics}}
  \textbf{\bibinfo{volume}{8}}, \bibinfo{pages}{14--16} (\bibinfo{year}{2012}).


\bibitem{barthelemy11}
\bibinfo{author}{Barth\'{e}lemy, M.}
\newblock \bibinfo{title}{Spatial networks}.
\newblock \emph{\bibinfo{journal}{Phys. Rep.}} \textbf{\bibinfo{volume}{499}},
  \bibinfo{pages}{1 -- 101} (\bibinfo{year}{2011}).

\bibitem{barthelemy03} \bibinfo{author}{Barth\'{e}lemy, M.}  \newblock
  \bibinfo{title}{Crossover from scale-free to spatial networks}.
  \newblock \emph{\bibinfo{journal}{EPL (Europhysics Letters)}}
  \textbf{\bibinfo{volume}{63}(6)}, \bibinfo{pages}{915}
  (\bibinfo{year}{2003}).


\bibitem{bullmore12}
\bibinfo{author}{Bullmore, E.} \& \bibinfo{author}{Sporns, O.}
\newblock \bibinfo{title}{The economy of brain network organization}.
\newblock \emph{\bibinfo{journal}{Nat Rev Neurosci}}
  \textbf{\bibinfo{volume}{13}}, \bibinfo{pages}{336--349}
  (\bibinfo{year}{2012}).

\bibitem{chen06}
\bibinfo{author}{Chen, B.~L.}, \bibinfo{author}{Hall, D.~H.} \&
  \bibinfo{author}{Chklovskii, D.~B.}
\newblock \bibinfo{title}{Wiring optimization can relate neuronal structure and
  function}.
\newblock \emph{\bibinfo{journal}{Proc Natl Acad Sci (USA)}}
  \textbf{\bibinfo{volume}{103}}, \bibinfo{pages}{4723--4728}
  (\bibinfo{year}{2006}).


\bibitem{watts98}
\bibinfo{author}{Watts, D.~J.} \& \bibinfo{author}{Strogatz, S.~H.}
\newblock \bibinfo{title}{Collective dynamics of small-world networks}.
\newblock \emph{\bibinfo{journal}{Nature}} \textbf{\bibinfo{volume}{393}},
  \bibinfo{pages}{440--442} (\bibinfo{year}{1998}).

\bibitem{latora03}
\bibinfo{author}{Latora, V.} \& \bibinfo{author}{Marchiori, M.}
\newblock \bibinfo{title}{Economic small world behaviour in weighted networks}.
\newblock \emph{\bibinfo{journal}{Eur Phys J B}} \textbf{\bibinfo{volume}{32}(2)},
  \bibinfo{pages}{249--263} (\bibinfo{year}{2003}).



\bibitem{perez07}
\bibinfo{author}{P\'erez-Escudero, A.} \& \bibinfo{author}{de~Polavieja, G.~G.}
\newblock \bibinfo{title}{Optimally wired subnetwork determines neuroanatomy of
  {\em Caenorhabditis elegans}}.
\newblock \emph{\bibinfo{journal}{Proc Natl Acad Sci (USA)}}
  \textbf{\bibinfo{volume}{104}}, \bibinfo{pages}{17180--17185}
  (\bibinfo{year}{2007}).

\bibitem{perez09}
\bibinfo{author}{P\'erez-Escudero, A.}, \bibinfo{author}{Rivera-Alba, M.} \&
  \bibinfo{author}{de~Polavieja, G.~G.}
\newblock \bibinfo{title}{Structure of deviations from optimality in biological
  systems}.
\newblock \emph{\bibinfo{journal}{Proc Natl Acad Sci (USA)}}
  \textbf{\bibinfo{volume}{106}}, \bibinfo{pages}{20544--20549}
  (\bibinfo{year}{2009}).

\bibitem{sporns11}
\bibinfo{author}{Sporns, O.}
\newblock \emph{\bibinfo{title}{Networks of the Brain}}
  (\bibinfo{publisher}{MIT Press}, \bibinfo{address}{Cambridge, MA},
  \bibinfo{year}{2011}).

\bibitem{kaiser06}
\bibinfo{author}{Kaiser, M.} \& \bibinfo{author}{Hilgetag, C.~C.}
\newblock \bibinfo{title}{Nonoptimal component placement, but short processing
  paths, due to long distance projections in neural systems}.
\newblock \emph{\bibinfo{journal}{PLoS Comput Biol}}
  \textbf{\bibinfo{volume}{2}}, \bibinfo{pages}{e95} (\bibinfo{year}{2006}).

\bibitem{towlson13} \bibinfo{author}{Towlson, E.K.}
  \bibinfo{author}{Vertes, P.E.}  \bibinfo{author}{Ahnert, S.E}
  \bibinfo{author}{Schafer, W.R.}  \bibinfo{author}{Bullmore, E.T.}
  \newblock\bibinfo{title}{The rich club of the {\em C. elegans}
    neuronal connectome}.  \newblock\emph{\bibinfo{journal}J
    Neurosci}, \textbf{\bibinfo{volume}{33(15)}},
  \bibinfo{pages}{6380-6387} (\bibinfo{year}{2013}).

\bibitem{varshney11}
\bibinfo{author}{Varshney, L.~R.}, \bibinfo{author}{Chen, B.~L.},
  \bibinfo{author}{Paniagua, E.}, \bibinfo{author}{Hall, D.~H.} \&
  \bibinfo{author}{Chklovskii, D.~B.}
\newblock \bibinfo{title}{Structural properties of the {\em Caenorhabditis elegans}
  neuronal network}.
\newblock \emph{\bibinfo{journal}{PLoS Comput Biol}}
  \textbf{\bibinfo{volume}{7(2)}}, \bibinfo{pages}{e1001066}
  (\bibinfo{year}{2011}).

\bibitem{kaiser11}
\bibinfo{author}{Varier, S.} \& \bibinfo{author}{Kaiser, M.}
\newblock \bibinfo{title}{Spatio-temporal development of the {\em
    Caenorhabditis elegans} neuronal network}.
\newblock \emph{\bibinfo{journal}{PLoS Comput Biol}}
\textbf{\bibinfo{volume}{7(1)}}, \bibinfo{pages}{e1001044}
(\bibinfo{year}{2011}).
  
  
\bibitem{arenas08}
\bibinfo{author}{Arenas, A.}, \bibinfo{author}{Fern\'{a}ndez, A.} \&
  \bibinfo{author}{G\'{o}mez, S.}
\newblock \bibinfo{title}{A complex network approach to the determination of
  functional groups in the neural system of {\em C. elegans}}.
\newblock In \bibinfo{editor}{Li\`{o}, P.}, \bibinfo{editor}{Yoneki, E.},
  \bibinfo{editor}{Crowcroft, J.} \& \bibinfo{editor}{Verma, D.~C.} (eds.)
  \emph{\bibinfo{booktitle}{Bio-Inspired Computing and Communication}},
  \bibinfo{pages}{9--18} (\bibinfo{publisher}{Lect. Notes Comp. Sci.
  Springer-Verlag}, \bibinfo{address}{Berlin, Heidelberg},
  \bibinfo{year}{2008}).

\bibitem{arenas09}
\bibinfo{author}{Arenas, A.}, \bibinfo{author}{Fern\'{a}ndez, A.} \&
  \bibinfo{author}{G\'{o}mez, S.}
\newblock \bibinfo{title}{11. an optimization approach to the structure of the
  neuronal layout of {\em C. elegans}}.
\newblock In \bibinfo{editor}{Boccaletti, S.} \& \bibinfo{editor}{Latora, V.}
  (eds.) \emph{\bibinfo{booktitle}{Handbook of biological networks}},
  vol.~\bibinfo{volume}{10}, \bibinfo{pages}{243--257}
  (\bibinfo{publisher}{World Scientific}, \bibinfo{address}{London,
    United Kingdom}
  \bibinfo{year}{2009}).


\bibitem{barabasi99}
\bibinfo{author}{Barabasi, A.-L.} \& \bibinfo{author}{Albert, R.}
\newblock \bibinfo{title}{Emergence of scaling in random networks}.
\newblock \emph{\bibinfo{journal}{Science}} \textbf{\bibinfo{volume}{286}},
  \bibinfo{pages}{509--512} (\bibinfo{year}{1999}).

\bibitem{mendes01}
\bibinfo{author}{Dorogovtsev, S.~N.} \& \bibinfo{author}{Mendes, J. F.~F.}
\newblock \bibinfo{title}{Effect of the accelerating growth of communications
  networks on their structure}.
\newblock \emph{\bibinfo{journal}{Phys. Rev. E}} \textbf{\bibinfo{volume}{63}},
  \bibinfo{pages}{025101} (\bibinfo{year}{2001}).


\bibitem{mckeown98}
\bibinfo{author}{McKeown, C.}, \bibinfo{author}{Praitis, V.} \&
  \bibinfo{author}{Austin, J.}
\newblock \bibinfo{title}{sma-1 encodes a $\beta_h$-spectrin homolog required
  for [\em {Caenorhabditis elegans} morphogenesis}.
\newblock \emph{\bibinfo{journal}{Development}} \textbf{\bibinfo{volume}{125}},
  \bibinfo{pages}{2087--2098} (\bibinfo{year}{1998}).

\bibitem{altun05} \bibinfo{author}{Altun, Z.~F.} \&
  \bibinfo{author}{Hall, D.~H.}  \newblock \bibinfo{title}{Handbook of
    {\em {C.} elegans} anatomy. {W}orm{A}tlas. Figure 6.} Available in
  the Introduction section
  at\\\url{www.wormatlas.org/hermaphrodite/hermaphroditehomepage.htm}
  (\bibinfo{year}{2012}).
  
 \bibitem{byerly76}
\bibinfo{author}{Byerly, L.}, \bibinfo{author}{Cassada, C.} \& \bibinfo{author}{Russel, R.~L.}
\newblock \bibinfo{title}{The life cycle of the nematode {\em {C}aenorhabditis elegans}}.
\newblock \emph{\bibinfo{journal}{Developmental Biology}} \textbf{\bibinfo{volume}{51}},
  \bibinfo{pages}{23-33} (\bibinfo{year}{1976}).
  
 
\bibitem{cuntz10}
\bibinfo{author}{Cuntz, H.}, \bibinfo{author}{Forstner, F.}, \bibinfo{author}{Borst, A.} \& \bibinfo{author}{H�usser, M.}
\newblock \bibinfo{title}{One rule to grow them all: A general theory of neuronal branching and its practical application}.
\newblock \emph{\bibinfo{journal}{PLoS Comput Biol}} \textbf{\bibinfo{volume}{6(8)}},
  \bibinfo{pages}{e1000877} (\bibinfo{year}{2010}). 

\bibitem{bassett10} \bibinfo{author}{Bassett, D.~S.},
  \bibinfo{author}{Greenfield, D.~L.},
  \bibinfo{author}{Meyer-Lindenberg, A.}, \bibinfo{author}{Weinberger,
    D.~R.},\bibinfo{author}{Moore, S.~W.} \&
  \bibinfo{author}{Bullmore, E.~T.}  \newblock
  \bibinfo{title}{Efficient physical embedding of topologically
    complex information processing networks in brains and computer
    circuits}.  \newblock \emph{\bibinfo{journal}{PLoS Comput Biol}}
  \textbf{\bibinfo{volume}{6}}, \bibinfo{pages}{ e1000748}
  (\bibinfo{year}{2010}).

\bibitem{heuvel09} \bibinfo{author}{van~den Heuvel, M.P.}
  \bibinfo{author}{Stam, C.J.}  \bibinfo{author}{Kahn, R.S.}
  \bibinfo{author}{Hulshoff Pol, H.E.}
  \newblock\bibinfo{title}{Efficiency of functional brain networks and
    intellectual performance}.\emph{\bibinfo{journal} J Neurosci}
  \textbf{\bibinfo{volume}{29}}, \bibinfo{pages}{7619-7624}
  (\bibinfo{year}{2009}).

\bibitem{fornito11}
\bibinfo{author}{Fornito, A.} \emph{et~al.}
\newblock \bibinfo{title}{Genetic influences on cost-efficient organization of
  human cortical functional networks}.
\newblock \emph{\bibinfo{journal}{J Neurosci}} \textbf{\bibinfo{volume}{31}},
  \bibinfo{pages}{3261--3270} (\bibinfo{year}{2011}).

\bibitem{vertes12}
\bibinfo{author}{V\'ertes, P.~E.} \emph{et~al.}
\newblock \bibinfo{title}{Simple models of human brain functional networks}.
\newblock \emph{\bibinfo{journal}{Proc Natl Acad Sci (USA)}}
  \textbf{\bibinfo{volume}{109}}, \bibinfo{pages}{5868--5873}
  (\bibinfo{year}{2012}).

\bibitem{vandenheuvel12}
\bibinfo{author}{van~den Heuvel, M.}, \bibinfo{author}{Kahn, R.~S.},
  \bibinfo{author}{Goni, J.} \& \bibinfo{author}{Sporns, O.}
\newblock \bibinfo{title}{High-cost, high-capacity backbone for global brain
  communication}.
\newblock \emph{\bibinfo{journal}{Proc Natl Acad Sci (USA)}}
  \textbf{\bibinfo{volume}{109}}, \bibinfo{pages}{11372--11377}
  (\bibinfo{year}{2012}).

\bibitem{erd60}
\bibinfo{author}{Erd\"os, P.} \& \bibinfo{author}{R\'enyi, A.}
\newblock \bibinfo{title}{On the evolution of random graphs}.
\newblock \emph{\bibinfo{journal}{Publ. Math. Inst. Hung. Acad. Sci.}}
  \textbf{\bibinfo{volume}{5}}, \bibinfo{pages}{17--61} (\bibinfo{year}{1960}).


\end{thebibliography}
\end{document}